\documentclass[pra,twocolumn]{revtex4}
\usepackage{graphicx,url}
\begin{document}

\title{Entanglement can enhance the distinguishability of 
entanglement-breaking channels}
\author{Massimiliano F. Sacchi}
\affiliation{QUIT, Unit\`a INFM and Dipartimento di Fisica 
``A. Volta'', Universit\`a di Pavia, I-27100 Pavia, Italy.} 
\date{\today}

\begin{abstract}
We show the rather counterintuitive result that entangled input states
can strictly enhance the distinguishability of two
entanglement-breaking channels.
\end{abstract}

\maketitle

%%%%%%%%%%%%%%%%%%%%%%%%%%%%%%%%%%%%%%%%%%%%%%%%%%%%
The class of entanglement-breaking channels---trace-preserving
completely positive maps for which the output state is always
separable---has been extensively studied
\cite{eb1,eb2,Hv,King1,Shor,King2,VDC,shir2}. More precisely, a
quantum channel ${\cal E}$ is called entanglement breaking if $({\cal
E} \otimes I)(\Gamma )$ is always separable, i.e., any entangled
density matrix $\Gamma $ is mapped to a separable one. The convex
structure of entanglement-breaking channels has been thoroughly
analyzed in Refs. \cite{eb1,eb2}. Moreover, the properties of such a
kind of channels have allowed to obtain a number of results for the
hard problem of additivity of capacity in quantum information theory
\cite{Hv,King1,Shor,King2,VDC,shir2,tcv,hw,shir1}.

Channels which break entanglement are particularly noisy in some
sense. In order to check if a channel is entanglement-breaking it is
sufficient to look at the separability of the output state
corresponding just to an input maximally entangled state \cite{eb1},
namely ${\cal E}$ is entanglement-breaking iff $({\cal E} \otimes
I)(|\beta \rangle \langle \beta |)$ is separable for $|\beta \rangle =
d^{-1/2} \sum _{j=0}^{d-1} |j \rangle \otimes |j \rangle $, $d$ being
the dimension of the Hilbert space. Another equivalent condition
\cite{eb1} is that the channel ${\cal E} $ can be written as 
\begin{eqnarray}
{\cal E}
(\rho )=\sum _k \langle \phi _k |\rho |\phi _k \rangle |\psi _k
\rangle \langle \psi _k |
\;,
\end{eqnarray} 
where $\{ |\phi _k \rangle \langle \phi _k |\}$ gives a positive
operator-valued measure (POVM), namely $\sum _k |\phi _k \rangle
\langle \phi _k |=I $ \cite{fin}.  The last formulation has an
immediate physical interpretation: an entanglement-breaking channel
can be simulated by a classical channel, in the sense that the sender
can make a measurement on the input state $\rho $ by means of a POVM
$\{ |\phi _k \rangle \langle \phi _k |\}$, and send the outcome $k$
via a classical channel to the receiver who then prepares an
agreed-upon pure state $|\psi _k \rangle $. For the above reason one
could think that entanglement---the peculiar trait of quantum
mechanics---may not be useful when one deals with
entanglement-breaking channels. In fact, entanglement breaking
channels have zero quantum capacity \cite{hw}.

\par In this report, however, we will show a situation in which the
use of entanglement can be relevant also for entanglement-breaking
channels, namely when one is asked to optimally discriminate two
entanglement-breaking channels, as in the quantum hypothesis testing
scenario \cite{hel}. What we mean is that an entangled input state can
{\em strictly} enhance the distinguishability of two given
entanglement-breaking channels.  We will make use of some recent
results \cite{discr} on the optimal discrimination of two given
quantum operations. In particular, a complete characterization of the
optimal input states to achieve the minimum-error probability has been
given for Pauli channels \cite{discr}, along with a
necessary and sufficient condition for which entanglement strictly
improves the discrimination. Such a condition is the following. 

\par Given with a priori probability $p_1$ and $p_2=1-p_1$ two Pauli
channels
\begin{eqnarray}
{\cal E}_i (\rho )= \sum_{\alpha =0}^3 q_i^{(\alpha )}\, \sigma
  _\alpha \,\rho \, \sigma _\alpha \;,\qquad {i=1,2,} 
\end{eqnarray}
where $\{\sigma _1\,,\sigma _2\,,\sigma _3 \}= \{ \sigma _x\,,\sigma
_y\,,\sigma _z\}$ denote the customary spin Pauli matrices,   
$\sigma _0 = I$, and $\sum _{\alpha =0}^3 q_i^ {(\alpha)}
 = 1$, the use of entanglement strictly improves the
discrimination iff \cite{discr}
\begin{eqnarray}
\Pi _{\alpha =0}^3 \, r_\alpha < 0 \;,\label{}
\end{eqnarray}
with 
\begin{eqnarray}
r_\alpha =p_1 \, q_1^{(\alpha )} - p_2 \, q_2^{(\alpha )}
\;.\label{ral}
\end{eqnarray}
Moreover, the optimal input state can always be chosen as a maximally
entangled state.

\par In the following we explicitly show the case of two
entanglement-breaking channels that are strictly better discriminated
by means of a maximally entangled input state. Let us consider for
simplicity two different depolarizing channels
\begin{eqnarray}
{\cal E}_i^ D(\rho )= q_i \, \rho + \frac {1-q_i}{3}\, 
\sum_{\alpha =1}^3 \sigma
  _\alpha \, \rho \, \sigma _\alpha \;, \qquad q_1 \neq q_2\;,
\end{eqnarray}
The two channels are supposed to be given with a priori probability
$p_1=p$ and $p_2=1-p$, respectively.  The coefficients $r_\alpha $ of
Eq. (\ref{ral}) are given in this case by 
\begin{eqnarray}
&& r_0=p\,q_1 -(1-p)\,q_2 \;, 
\nonumber \\& & 
r_1=r_2=r_3=p \, \frac {1-q_1}{3}- (1-p)\,\frac
  {1-q_2}{3}\;.
\end{eqnarray}
Hence, entanglement strictly enhances the distinguishability of the
two channels ${\cal E}_1^D$ and ${\cal E}_2^D$  
iff 
\begin{eqnarray}
&&[p\,q_1 -(1-p)\,q_2] 
\left [ p \,\frac {1-q_1}{3}- (1-p)\,\frac
  {1-q_2}{3}\right ] 
< 0\;,\label{cond1}
\end{eqnarray}
or equivalently 
\begin{eqnarray}
&&(q_1+q_2)(2-q_1-q_2) p^2 -(q_1-2q_1q_2+3q_2-2q_2^2)p \nonumber \\& &
  + q_2(1-q_2)<0 
%(q_1+q_2)(2-q_1-q_2) \, p^2 -(q_1-2q_1q_2+3q_2-2q_2^2)\, p  + q_2(1-q_2)<0 
\;.\label{cond2}
\end{eqnarray}
The solution of Eq. (\ref{cond2}) for the prior probability $p$
versus $q_1$ and $q_2$ is given by 
\begin{eqnarray}
&&\frac{1-q_2}{2-q_1-q_2} < p < \frac{q_2}{q_1+q_2}\qquad \hbox{for
}\quad q_1 < q_2 \;, \nonumber \\& & 
\frac{q_2}{q_1+q_2}< p < \frac{1-q_2}{2-q_1-q_2} 
\qquad \hbox{for
}\quad q_1 >  q_2
\;.\label{sol}
\end{eqnarray}
A depolarizing channel is entanglement breaking iff $q \leq 1/2$,
where $q$ is the probability pertaining to the identity
transformation. This fact can be easily checked by applying the PPT
condition \cite{ppt1,ppt2} to the Werner state \cite{ws} $({\cal
E}\otimes I)(|\beta \rangle \langle \beta |)$, where $|\beta \rangle $
denotes the maximally entangled state $|\beta \rangle =\frac{1}{\sqrt
2}(|00 \rangle +|11 \rangle )$. It follows that the solution in
Eq. (\ref{sol}) for $q_1,q_2 \leq 1/2$ gives examples of situations
where a maximally entangled input state strictly improves the
distinguishability of two entanglement-breaking channels.  

\begin{figure}[htb]
\begin{center}
\includegraphics[scale=1]{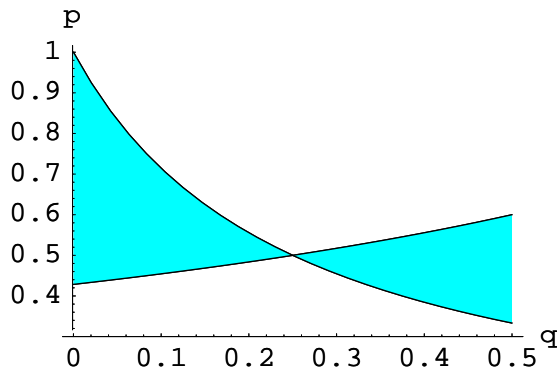}
\caption{The grey region represents the value of the a priori
probability $p$ for which the discrimination between a depolarizing
channel with $q\leq 1/2$ (an entanglement-breaking channel) and a
completely depolarizing channel is strictly enhanced by using a
maximally entangled input state.}
\label{f:fig1}
\end{center}
\end{figure}

\par In Fig. 1 we plot such a set of solutions for the a priori
probability $p$ in the case of discrimination between an
entanglement-breaking depolarizing channel with $q_1=q\leq 1/2$ and a
completely depolarizing channel $q_2=1/4$.

\par In conclusion, in the problem of discriminating two quantum
operations the relevant object is the map corresponding to the their
difference, which is not a completely positive map. Using entangled
states at the input of entanglement-breaking channels give output
separable states that, however, can be better discriminated since they
live in a higher dimensional Hilbert space.  Curiously, we note that,
on the other hand, when we are asked to optimally discriminate two
arbitrary unitary transformations---which are of course
entanglement-preserving operations---entanglement never enhances the
distinguishability \cite{1,2,3}.

\emph{Acknowledgments.} This work has been sponsored by INFM through
the project PRA-2002-CLON, and by EC and MIUR through the cosponsored
ATESIT project IST-2000-29681 and Cofinanziamento 2003.


\begin{thebibliography}{99}
\bibitem{fin}For simplicity,  we are considering finite dimensional
  Hilbert space, for which the POVM has rank-one elements. This may
  not be the case in infinite dimension \cite{shir2}.
\end{thebibliography}
\end{document}